\documentclass[conference]{IEEEtran}
\ifCLASSINFOpdf
\else
\fi
\newcommand\Mark[1]{\textsuperscript{#1}}

\usepackage{diagbox} %table split headers
\usepackage{algpseudocode}
\usepackage{caption}
\usepackage{subcaption}
\usepackage{graphicx}
\usepackage{epstopdf}
\usepackage{booktabs}
\usepackage{tabularx}
\usepackage{multirow}
\usepackage{float}
\usepackage{color}
\usepackage{tabu}
\usepackage{comment}
\usepackage{dcolumn}
\usepackage{amsmath}
\usepackage {CJK}
\usepackage[normalem]{ulem}
\usepackage[ruled, linesnumbered]{algorithm2e} %ruled,linenumbered
\usepackage{bm}
\usepackage{mathtools}
\usepackage{arydshln}

\makeatletter
\newcases{nocases}
{\quad}
{$\m@th\displaystyle{##}$\hfil}
{$\m@th\displaystyle{##}$\hfil}
{.}{.}
\makeatother

\begin{document}
	
	\title{Fast and Reliable WiFi Fingerprint Collection for Indoor Localization}
		\author{\IEEEauthorblockN{Fuqiang Gu, Milad Ramezani, Kourosh Khoshelham}
			\IEEEauthorblockA{ Department
				of Infrastructure Engineering,\\
				University of Melbourne\\
				Emails: fuqiangg@student.unimelb.edu.au,\\
				mramezani@student.unimelb.edu.au,\\
				k.khoshelham@unimelb.edu.au}
			\and 
			\IEEEauthorblockN{Xiaoping Zheng, Ruqin Zhou, Jianga Shang}
			\IEEEauthorblockA{\Mark{1} Faculty of Information Engineering, \\
			  China University of Geosciences;\\
			  \Mark{2} National Engineering Research Center \\for Geographic Information System \\
				Emails: (zheng.xiaoping, zhouruqin, jgshang)@cug.edu.cn}
			}


	% for Transactions on Magnetics papers, we must declare the abstract and
	% index terms PRIOR to the title within the \IEEEtitleabstractindextext
	% IEEEtran command as these need to go into the title area created by
	 \maketitle
%	\IEEEpeerreviewmaketitle
	%\IEEEtitleabstractindextext{%
	\begin{abstract}
	Fingerprinting is a popular indoor localization technique since it can utilize existing infrastructures (e.g., access points). However, its site survey  process is a labor-intensive and time-consuming task, which limits the application of such systems in practice. In this paper, motivated by the availability of advanced sensing capabilities in smartphones, we propose a fast and reliable fingerprint collection method to reduce the time and labor required for site survey. The proposed method uses a landmark graph-based method to automatically associate the collected fingerprints, which does not require active user participation. We will show that besides fast fingerprint data collection, the proposed method results in accurate location estimate compared to the state-of-the-art methods. Experimental results show that the proposed method is an order of magnitude faster than the manual fingerprint collection method, and using the radio map generated by our method achieves a much better accuracy compared to the existing methods.
		
	\end{abstract}
	
	%}

	\IEEEdisplaynontitleabstractindextext

	% \fi
	\IEEEpeerreviewmaketitle

\section{Introduction}
Indoor localization is important for a number of applications such as shopping guide, augmented reality game, mobile social networks, and location-based services. WiFi-based methods are popular because of the ubiquity of WiFi infrastructure (e.g., access points) and their low cost \cite{1,2}. WiFi-based localization methods can be categorized as: model-based and fingerprint-based. Model-based methods rely on a signal propagation model, which converts the received signal strength (RSS) from an access point (AP) to the distance between the receiver (user's device) and the AP. The location of the user can be computed by trilateration after obtaining more than three distances to different APs. However, model-based methods require the knowledge of APs' locations, which are difficult to obtain in many cases. Besides, it is challenging to build a precise signal propagation model due to multipath effect, shadowing, fading, and delay distortion. An inaccurate signal propagation model will lead to a large localization error when using trilateration. 

In order to overcome the limitations of model-based methods, fingerprint-based methods have been proposed. Compared with model-based methods, fingerprint-based methods do not require the knowledge of APs' locations and signal propagation models, and can achieve a higher localization accuracy. Fingerprint-based localization includes two phases: offline training and online localization. In the offline training phase, a site survey is conducted to collect fingerprints at known locations called reference points (RPs). Each fingerprint contains the Media Access Control (MAC) address of visible APs and their corresponding RSS. These fingerprints together with their respective RPs' coordinate are then stored in a radio map, which may also be called the heat map. In the online localization phase, the newly-collected fingerprint is matched with the most similar fingerprints stored in the radio map, from which the user's location can be obtained.

The main challenge of WiFi fingerprinting is the construction of a fingerprint database (also known as radio map), which is a labor-intensive and time-consuming process. Many research works have been done to expedite the site survey process \cite{3,4}. One of the popular methods is RSS prediction by interpolation or by signal propagation modeling \cite{5}. WiFi SLAM (short for Simultaneous Localization and Mapping) \cite{6,7,8,9} is another popular way to construct the radio map with low survey cost, but its heavy computational load makes it unsuitable to work on the resource-limited handheld devices such as smartphones. The crowdsourcing-based radio map construction has been studied recently, including active crowdsourcing \cite{10,11,12} and passive crowdsourcing \cite{13,14,15}. The active crowdsourcing methods use intrusive/explicit user feedback to build a radio map. Although active crowdsourcing eliminates the need for professional surveyors, it requires active user participation and may suffer from intentional frauds. To reduce the active user participation, the passive fingerprint crowdsourcing has been proposed. Passive crowdsourcing associates fingerprints to corresponding RPs with the aid of smartphone inertial sensors. While the systems using passive crowdsourcing have made fingerprinting more practical than before, they still have various limitations such as the need for GPS, limited applicability, and low accuracy. 
 
Although a lot of efforts have been made to reduce the labor and time required to construct a radio map, they suffer from various limitations such as requirement for active user participation, being computationally expensive, limited applicability, and low accuracy. In this study, we propose a novel fast WiFi fingerprint collection method, which uses a landmark graph-based localization method for automatically estimating the location of RPs associated with the collected fingerprints. The landmark graph-based method has been shown to achieve a better localization accuracy than map filtering and pedestrian dead reckoning (PDR) \cite{16}. To ensure the accuracy of the constructed radio map, we design a metric (belief) to evaluate the quality of location estimation of RPs, and only the location estimates with a high ``belief'' are used to associate corresponding fingerprints. Compared to existing fingerprint collection methods, our method is able to construct a more accurate and reliable radio map at a very low cost, enabling WiFi-based localization with a higher accuracy.

\section{System Overview}
\begin{figure}[h]
	\centering 
	\includegraphics[width=0.4\textwidth]{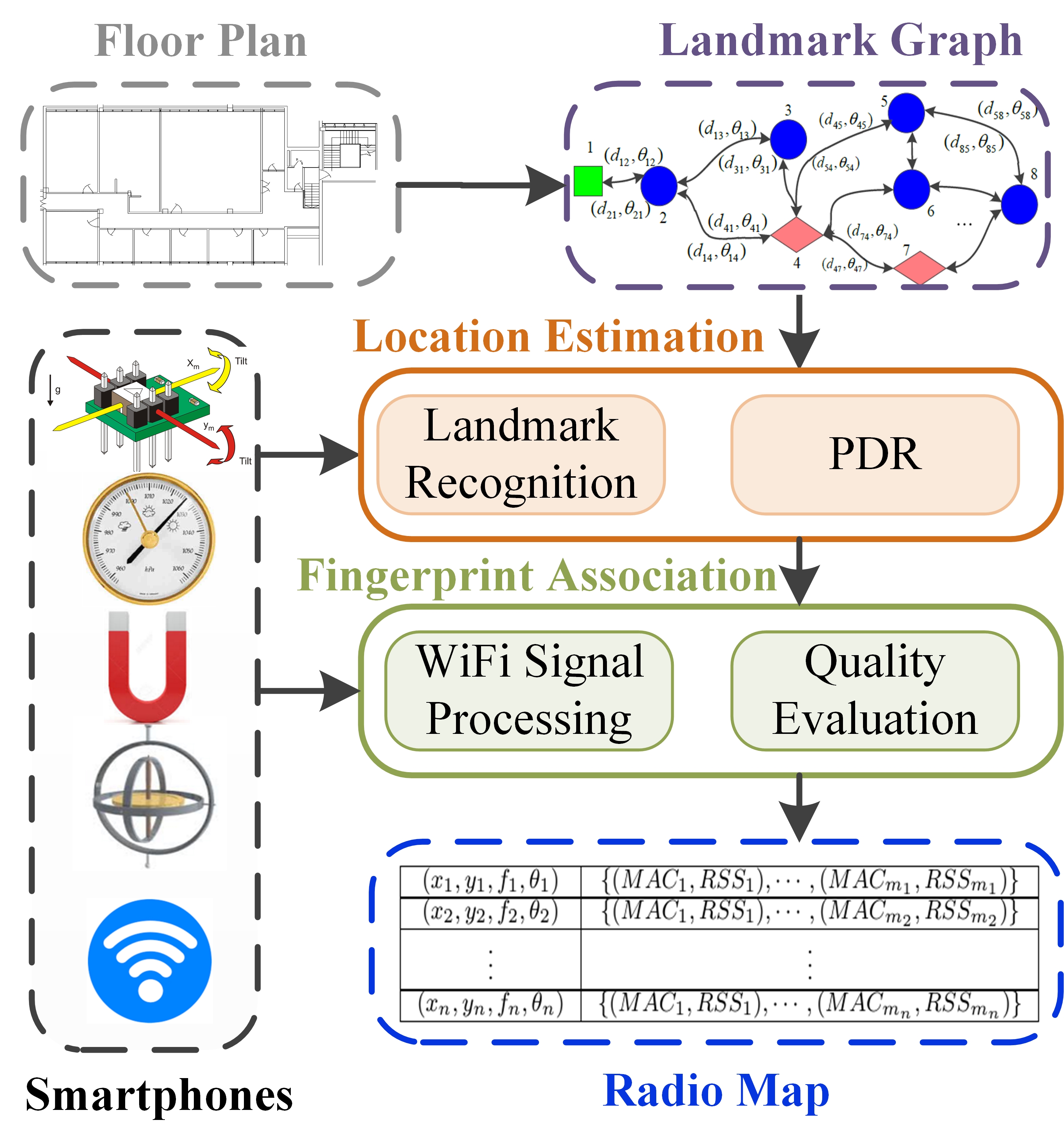}
	\caption{System architecture}
	\label{fig:architecture}
\end{figure}

As shown in Figure \ref{fig:architecture}, the system architecture encompasses two main modules: Location Estimation of RPs, and Fingerprint Association. The Location Estimation module takes as input a landmark graph \cite{16}, and the readings from the accelerometer, magnetometer, gyroscope, and barometer built in today's smartphones. These readings are used to compute the user's step length and heading, and infer her location in real time by using the PDR method \cite{23}. The accumulative error of PDR is bounded by using the landmark graph. When a landmark from the landmark graph is detected in sensor data, the user's location will be calibrated by using the location of the detected landmark. 
The Fingerprint Association module uses the location estimated by the Location Estimation module, and the scanned WiFi signal information (including the MAC address and corresponding RSS between the smartphone and each AP) to generate the radio map. To guarantee the accuracy and reliability of the radio map, we evaluate the quality of estimated location from the Location Estimation module. Only the location estimates that meet the quality requirement are used to associate the corresponding fingerprint. Such location estimate together with the corresponding fingerprint are then added into the radio map.

\section{Landmark Graph-based Location Estimation}
\subsection{Landmark Detection}
Landmarks in this research refer to location points where sensor data present a distinct, stable change pattern. Corners or turns, for instance, compel users to change their walking direction; a door imposes users to switch their motion state from Walking to Still. The type of landmarks of our interest includes Accelerometer landmarks, Gyroscope landmarks, and Barometer landmarks, which will be discussed in the following. The locations of these landmarks correspond to the locations of doors, elevators, stairs, corners and turns that can be simply obtained from floor plans or quickly measured. Mathematically, we define a landmark $v$ as follows:
$$
v=<(x,y,f),(\mathcal{R}_1,\cdots,\mathcal{R}_M)> \eqno (1)
$$ 
where $(x,y,f)$ denotes the location of the landmark, $(\mathcal{R}_1, \cdots, \mathcal{R}_M)$ represents the detection rule in different types of sensor readings, $M$ is the number of rules that this landmark possesses. A landmark may satisfy one or more types of detection rules.  
\begin{figure}[ht]
	\centering
	\includegraphics[width=0.4\textwidth]{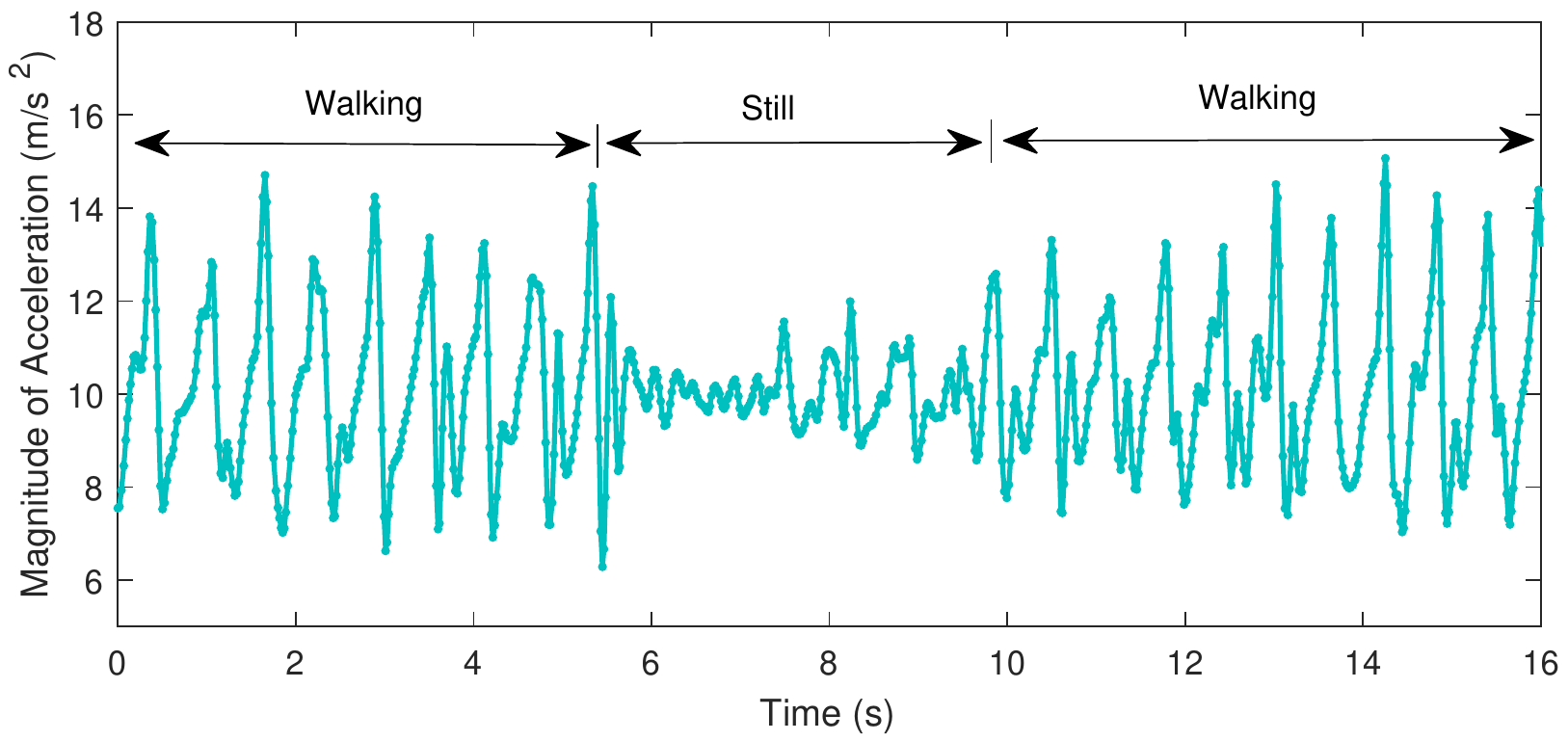}
	\caption{The acceleration changes as the user passes a door}
	\label{fig:acc_landmark}
\end{figure}

\textbf{Accelerometer landmark} --  Accelerometer landmarks refer to location points where the motion state of the user presents a distinct change pattern, which can be sensed by the accelerometer. A location point that witnesses the change pattern of ``Walking $\longrightarrow$ Still (for a few seconds) $\longrightarrow$ Walking" can be regarded as a potential accelerometer landmark. This pattern may arise when the user passes a door (as shown in Figure \ref{fig:acc_landmark}), which can be detected by thresholding the magnitude of acceleration. A location point is regarded as an Accelerometer landmark if the accelerometer readings present this change pattern every time the user passes it. Mathematically, the rule $\mathcal{R}_{acc}$ of Accelerometer landmarks is defined as:
\begin{equation}
\begin{nocases}
\mathcal{R}_{acc} &= (loc_t|m_{t-K_1:t} == \text{walking}\ \nonumber \\
& \qquad \&\& \ m_{t:t+K_2} == \text{still} \nonumber \\ 
& \qquad \&\& m_{t+K_2:t+K_2+K_1} == \text{walking})  \nonumber 
\end{nocases} 
\eqno (2)
\end{equation}
where $m_t$ represents the user's motion state (e.g, walking, still) at time $t$. $K_1$ and $K_2$ are two thresholds that determine the period of the corresponding motion state which can be empirically set.

\begin{figure}[ht]
	\centering
	\includegraphics[width=0.4\textwidth]{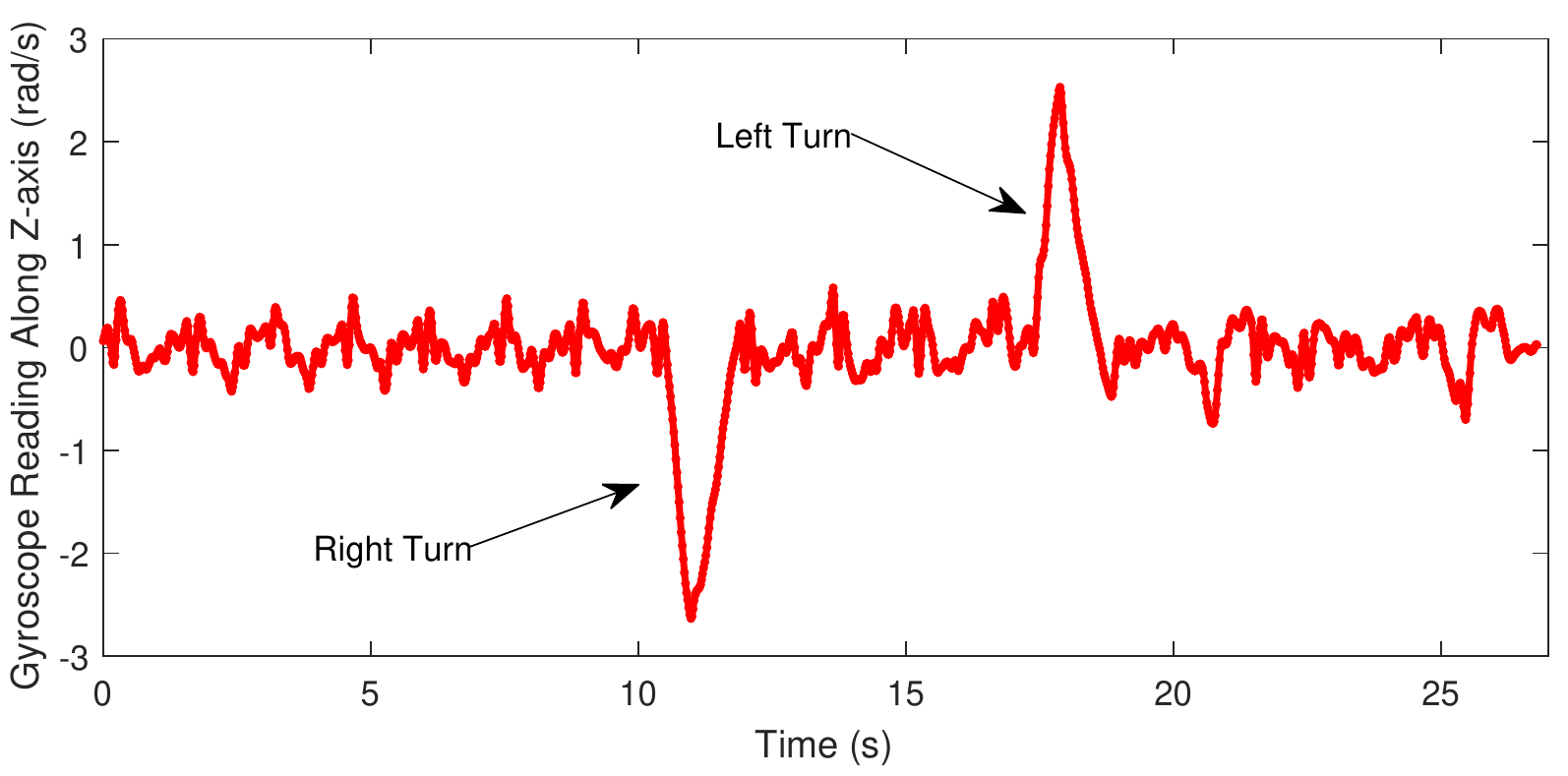}
	\caption{The gyrocope readings change as the user takes a right turn or a left turn}
	\label{fig:gyro_landmark}
\end{figure}

\textbf{Gyroscope landmark} -- A Gyroscope landmark is defined as a location point where the gyroscope readings present a distinct and stable pattern. Although the magnetometer can be also used to detect the change of walking direction, its readings tend to be easily affected by ferromagnetic materials. Therefore, we use the gyroscope readings to detect the change in the walking direction of the user. Figure \ref{fig:gyro_landmark} shows how the gyroscope readings change when the user takes a right or left turn. A Gyroscope landmark is usually witnessed at the location of a turn, corner or door. To detect a gyroscope landmark, we define the rule $\mathcal{R}_{gyro}$ of detecting a gyroscope landmark as follows:
\[ \mathcal{R}_{gyro}= (loc_t|\left|\dot{\theta_t} \right| > \epsilon_{gyro})
\eqno (3)
\]
where $\dot{\theta}_t$ is the gyroscope readings along the vertical direction. If the absolute value of $\dot{\theta}_t$ is greater than a certain threshold $\epsilon_{gyro}$, we consider this location point as a potential Gyroscope landmark. 

\begin{figure}[H]
	\centering 
	\includegraphics[width=0.4\textwidth]{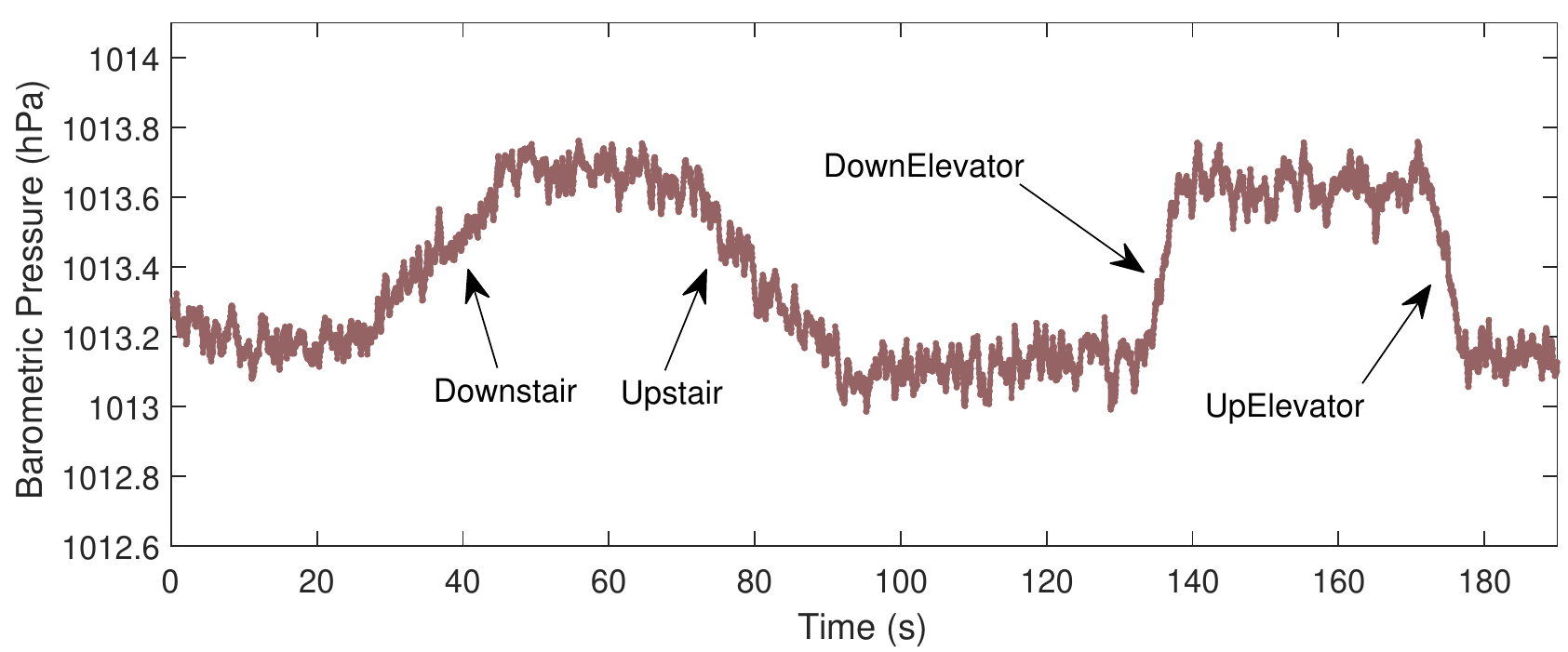}
	\caption{The change in the barometer readings when going stairs and taking an elevator}
	\label{fig:LM_baro}
\end{figure}

\textbf{Barometer landmark} --  The barometer is capable of detecting the vertical movement of a user (e.g., going upstairs or downstairs, taking an elevator) since the barometric value changes with the altitude or height. Although the barometric pressure is influenced by many factors such as temperature and altitude, we can argue that it is just affected by the altitude during a short period of time. 
Fig. \ref{fig:LM_baro} shows the change in the barometer readings when a user walks horizontally, goes upstairs/ downstairs, and takes an elevator upward/downward. The entrance and exit of stairs and elevators can be regarded as barometer landmarks since they present a pattern in barometer readings that is identifiable, distinctive, and stable. The entrance detection is done by detecting the change pattern ``horizontal movement $\longrightarrow$ vertical movement''. Similarly, the exit is detected by using the pattern ``vertical movement $\longrightarrow$ horizontal movement''. Both change patterns are recognized by utilizing the barometer readings. 

Let $p_i$ denote the average value of the $i$-th window of air pressure readings that contains the air pressure value at time $t$, and let $\epsilon_{baro_1}$ and $\epsilon_{baro_2}$ be the thresholds used to detect the user's horizontal movement and vertical movement, respectively. The rule to detect the entrance to a set of staircases or an elevator is defined as:
\begin{equation}
\begin{nocases}
\mathcal{R}_{baro_1} &= (loc_t| (\left|p_i-p_{i-1}\right|) < \epsilon_{baro_1}  \\
& \&\& \left|\sum_{j=i+1}^{i+K_{p_1}}(sgn(p_{j}-p_{j-1}))\right|==K_{p_1} \nonumber  \\ 
& \&\& |p_{i+K_{p_1}}- p_i|> \epsilon_{baro_2}) \nonumber  \\
\end{nocases} 
.
\eqno (4)
\end{equation}
The first term is for detecting horizontal movement, and the latter two terms are for detecting vertical movement. $sgn$ is the sign function, which is described as:
$$
sgn(p_{j}-p_{j-1}) = \left \{ 
\begin{array}{llll}
1, \qquad \text{if $p_{j} > p_{j-1}$}\\
0, \qquad \text{if $p_{j} = p_{j-1}$}\\
-1, \ \quad \text{if $p_{j} < p_{j-1}$}\\
\end{array} 
\right. 
.
\eqno (5)
$$
Similarly, we can define the rule to detect the exit from a set of staircases or an elevator as:
\begin{equation}
\begin{nocases}
\mathcal{R}_{baro_2} &= (loc_t|(|p_i-p_{i+1}|) < \epsilon_{baro_1}  \\
& \&\& \left|\sum_{j=i-K_{p_2}+1}^{i}(sgn(p_{j}-p_{j-1}))\right|==K_{p_2} \nonumber  \\ 
& \&\& |p_{i-K_{p_2}}- p_i|> \epsilon_{baro_2}) \nonumber  \\
\end{nocases} 
.
\eqno (6)
\end{equation}
The values of $K_{p_1}$ and $K_{p_2}$ are not constant, but are determined dynamically. Their initial values are set to 1, and gradually increase as long as the value of the sign function keeps unchanged. 

\subsection{Landmark Graph for Estimating the Location of RPs}
To accurately estimate the location of RPs, we adopt the landmark graph-based method \cite{16}, which uses the powerful sensing ability of smartphone sensors to the landmarks that are naturally distributed in indoor environments. By using the smartphone sensors to detect these landmarks, the accumulative error of PDR can be bounded, which results in an accurate location estimation of RPs. 

A landmark graph is a directed graph where nodes are landmarks and edges are accessible paths with heading information. Let $G=(V,E)$ denote a landmark graph where $V=\{v_1,\cdots, v_N\}$ is a set of landmarks and $E=\{e_1,\cdots,e_M\}$ is the set of edges in graph $G$. Each edge $e_i=<v_j,v_k,\theta_{jk},d_i>$ is a tuple consisting of two landmarks, the direction from one landmark to another, and the distance between these two landmarks. Note that the direction from landmark $v_j$ to landmark $v_k$ is different from that from $v_k$ to landmark $v_j$, which means that there are two edges between two neighboring landmarks. The complete algorithm of using a landmark graph for indoor localization consists of the following steps:  

1) \textbf{Constructing the landmark graph of the indoor environment}. The construction of a landmark graph requires the location landmarks, which can be extracted from a floor plan. Since most buildings are symmetric, which means that different floors for the same building have similar layout, a landmark graph for one floor can be easily modified for another floor by simply changing the floor information in the landmarks' coordinates. Therefore, even manually constructing a landmark graph is not onerous and needs to be done once only.     

2) \textbf{Detecting step events using the accelerometer readings}. The step detection is done by detecting acceleration peaks. Each peak corresponds to one step if the variance of the acceleration within the detected time window is greater than 0.5 $m/s^2$, which is a threshold to distinguish Walking state from Still state. If a step is detected, then we compute the corresponding heading and step length. 

3) \textbf{Estimating the location of the user at each step using PDR}. Given an initial location, we can apply the PDR algorithm to compute the user's location in real time using the inertial sensor readings. Let $(x_t, y_t, f_t)$ denote the location of a user at time $t$ where $(x_t, y_t)$ is the coordinate and $f_t$ is the floor on which the user is, $l_t$ is the corresponding step length, and $\theta_t$ represents the heading. Then the location estimation using the PDR method can be described as:
$$
x_t=x_{t-1} + l_t cos(\theta_t)  \eqno (7) \label{eq:pdr_x} 
$$
$$
y_t=y_{t-1} + l_t sin(\theta_t)  \eqno (8) \label{eq:pdr_y}
$$
$$
f_t = f_{t-1} -  \frac{p_t - p_{t-1}}{p_d}  \eqno (9) 
$$
where $p_t$ and $p_{t-1}$ are the air pressure values at time $t$ and $t-1$, respectively. $p_d$ is the air pressure difference between two floors. %, which is about 0.45 hPa for our test environments. 

4) \textbf{Detecting landmarks and correcting the estimated location}. While the PDR is conducted to estimate the location of the user, the readings from accelerometer, gyroscope, and barometer are simultaneously used to detect landmarks encountered. Although both the gyroscope readings and compass readings (inferred from accelerometer readings and magnetometer readings) can be used to estimate the heading, they cannot provide a robust heading estimation since the gyroscope has the drift problem and the compass is vulnerable to ferromagnetic materials. Therefore, we use the landmark graph to assist with the heading estimation. If the user is detected to walk on the path connecting two landmarks in the landmark graph, the heading from this landmark graph will be used. Otherwise, the compass readings will be used as the heading. Also, the user's step length is updated when she passes two neighboring landmarks in the landmark graph. Let $v_1$ and $v_2$ indicate the two neighboring landmarks that a user passes subsequently, then we can calculate the step length $l$ as follows:
$$
l= \frac{dist(v_1,v_2)}{N_s} \eqno (10)
$$
where $dist(v_1,v_2)$ is the Euclidean distance between $v_1$ and $v_2$, and $N_s$ is the number of detected steps. The advantage of this step length estimation method is that it does not require the user's stature information and can adapt to varying walking speeds since it is updated as the user passes two neighboring landmarks on the landmark graph.

A key challenge to using landmarks for assisting localization is to solve the data association issue. In other words, when there are multiple landmarks nearby, it is difficult to determine the detected landmark. It is also challenging to deal with the case that one or more landmarks are missed in sensor data. To solve the above challenges, we define a measure $con$ to indicate how much confidence we place on the location point meeting the landmark detection rule. The confidence that the location point $(x_t,y_t)$ is matched with the landmark $v_k$ in the landmark graph is expressed as:
$$
con(v_k) = \delta (R_k, R_t^*) \cdot r (\theta_k,\theta_t^*) \cdot g(d_k,d_t^*) \eqno (11)
$$
where $k$ is the index of landmark in the landmark graph. $R_k$ the detection rule of the reference landmark $v_k$, and $R_t*$ is the type of the detected landmark at time $t$. $\theta_k$ and $\theta_t^*$ are the reference heading and the estimated heading from the time visiting the last landmark to time $t$. $d_k$ and $d_t^*$ are the reference distance and the traveled distance from last landmark to location $(x_t,y_t)$. $\delta$ is the Dirac delta function, which is defined as
$$
\delta (R_k, R_t^*) = \left \{ 
\begin{array}{llll}
1, \qquad \text{if $R_k$ == $R_t^*$}\\
0, \qquad \text{otherwise}\\
\end{array} 
\right.
,
\eqno (12)
$$
and $r$ is the rectangle function, described as 
$$
r (\theta_k,\theta_t^*) = \left \{ 
\begin{array}{llll}
1, \qquad \text{if $|\theta_k - \theta_t^*| < \epsilon_{theta}$}\\
0, \qquad \text{otherwise}\\
\end{array} 
\right.
\eqno (13)
$$
where $\epsilon_{theta}$ is a heading threshold. The function $g$ is defined as
$$
g(d_k,d_t^*)  = 1/\left|d_k -d_t* \right|.  \eqno (14)
$$

If there are multiple possible landmarks nearby, the one with the largest confidence will be chosen as the detected landmark. To reduce the risk of mis-matching, we set a confidence threshold $\epsilon_{v}$ to exclude fake landmarks. For instance, a user may take a turn in the middle of a corridor, which may be mistakenly detected as a gyroscope landmark. However, since fake landmarks have a lower confidence value, we are able to exclude them by judging whether their confidence is smaller than $\epsilon_v$. We use only the landmarks with confidence larger than $\epsilon_v$ to calibrate the accumulative error. 

In some cases, some landmarks may be missed in sensor data. For example, certain landmarks at the locations of doors will be missed if a door is left open since the user does not behave in the ``Walking $\longrightarrow$ Still $\longrightarrow$ Walking" pattern. In these cases, we simply ignore them and do not calibrate the user's location until next landmark is detected.

\section{Construction of WiFi Radio Map}

\subsection{Quality Evaluation of Estimated Location}
To construct an accurate and robust radio map, we evaluate the quality of estimated location before using it to associate a fingerprint. This will ensure that only accurate location estimates are included in the radio map.% We believe that the accurate location estimation of RPs will lead to generating an accurate radio map.

The PDR method consists of two components: step length estimation, and heading estimation. Therefore, the localization accuracy of PDR depends on the accuracy of step length estimation and heading estimation. Since we use the landmark graph to bound the accumulative error of PDR, we are able to achieve an accurate location estimation when the user walks normally at a relatively stable walking speed. Thus, the error of the landmark graph-based PDR method mainly comes from varying walking speeds, false walking (which means the user remains still while using her smartphone for texting, call, or playing games), and frequent stops. It has been shown in \cite{24} that the step periodicity for the same motion state (e.g., walking, jogging) exhibits little variation when the user moves at a relatively constant speed. However, the periodicity varies significantly when the user remains still while using the phone arbitrarily, e.g., for texting, playing phone games, etc. By limiting the range of step periodicity to a certain interval, we are able to reduce the location estimation error of RPs. The step periodicity is defined as the time period of taking one step (one left step or right step), which equals the time interval between two neighboring peaks of the amplitude of accelerometer readings. 

Based on the periodicity of the steps, we define a metric $bel$ to measure the quality of the location estimation of RPs. Let $S$ denote a path segment connecting two landmarks that the user takes $N$ steps to travel, including $N+1$ RPs and the corresponding time $t_i$ when the user visited it, namely 
$$
S= \{(t_1, x_1,y_1,f_1),\cdots, (t_{N+1}, x_{N+1},y_{N+1},f_{N+1})\}. \eqno(15)
$$ 
Let $T$ indicate the step periodicity set that contains the step periodicity of these $N$ steps, namely
$$
T= \{T_1,\cdots, T_N\} \eqno(16)
$$ 
where $T_i$ is period of the $i$-th step measured by the accelerometer. Thus, the metric to measure the quality of the path segment $S$ is defined as:
$$
bel(S) = \frac{\sum_{i=1}^N \chi(T_i)\cdot T_i}{\sum_{i=1}^N T_i} \cdot \frac{1}{\sigma_{T'}} \eqno(17)
$$
where the first term indicates the ratio of valid steps that exclude outliers caused by false walking or frequent stops. $\chi$ is the identifier function, which is defined as 
$$
\chi (T_i) = \left \{ 
\begin{array}{llll}
1, \qquad \text{if $T_i \in [T_{min}, T_{max}]$}\\
0, \qquad \text{otherwise}\\
\end{array} 
\right.
.\eqno (18)
$$
$T'$ is a subset of $T$, representing the step periodicity set of valid steps and consisting of elements of $T$ that fall in the valid interval $[T_{min}, T_{max}]$. If a user walks at a stable speed, then the deviation of her step periodicity vector $T$ should be small. To enable the metric to reflect this stability, we take the reciprocal of the standard deviation $\sigma_{T'}$ of $T'$ into the metric. When the belief $bel(S)$ of the path segment $S$ is greater than a threshold $\epsilon_S$, the location estimation on this path segment is considered accurate and reliable.

\subsection{Construction of WiFi Radio Map}

Let $Tr$ denote a trajectory that the user has traveled, including $K$ path segments that are divided by landmarks, namely $Tr =\{S_1, \cdots, S_K \}$, and $FP$ is a set of fingerprints collected along this trajectory, $FP = \{fp_1, \cdots, fp_N\}$ where $N$ is the number of WiFi scans. Since the time when a step event happens may be different from the time when a WiFi scan is conducted, we need to synchronize the time to conduct a WiFi scan with the time a step event happens in order to associate the fingerprint with the estimated location. Suppose that the WiFi scan at time $t_j$ happens during the time period the user walks from location $(x_{k-1}, y_{k-1})$ to $(x_k,y_k)$, namely $t_{k-1} \leq t_j \leq t_k$, then we can estimate the location of the $j$-th potential RP by using a linear interpolation, namely
$$
x_j =  x_{k-1} + \frac{(x_k-x_{k-1})\cdot (t_j - t_{k-1})}{t_k-t_{k-1} }
\eqno (19)
$$

$$
y_j = y_{k-1} + \frac{(y_k-y_{k-1})\cdot (t_j - t_{k-1})}{t_k-t_{k-1}}.
\eqno (20)
$$

The complete procedure of constructing a radio map is described in Algorithm \ref{alg:construction_radio_map}. We first use the linear interpolation method to obtain the location $(x_j, y_j)$ of the $j$-th potential RP. According to the calculated location $(x_j, y_j)$, we can find the path segment $S_i$ that includes the location $(x_j, y_j,f_j)$. After this, we evaluate the quality of $S_i$. Only when the $bel(S_i)$ meets the defined requirement, the fingerprint is associated with the corresponding location and then is added to the radio map. This process is recursively done until all the elements in the $FP$ set are used. %The constructed radio map $rm$ is as follows:

\begin{algorithm}[h]
	\caption{Radio map construction}
	\label{alg:construction_radio_map}
	\SetKwInOut{Input}{Input}
	\SetKwInOut{Output}{Output}
	\Input{A trajectory $Tr =\{S_1, \cdots, S_K \}$, a set of RSS vector $FP = \{fp_1, \cdots, fp_N\}$} 
	\Output{A radio map}
	% Calculate the value of $RSS_0$, $n_p$, and $\sigma_{RSS}$ for each AP \\
	\For{$j = 1:N$}{
		Compute the location ($x_j,y_j,f_j$) of the $j$-th potential RP;\\
	%	Obtain the user' heading $\theta_j$ at time $t_j$;\\
		Search for the segment $S_i$ that includes location ($x_j,y_j,f_j)$; \\
		Compute the belief $bel(S_i)$ of the segment $S_i$;\\
	%	Compute the belief $bel(RSS_{t_j})$ of the RSS at time $t_j$;\\
		\If{$bel(S_i) > \epsilon_S$}{
			Associate the fingerprint $fp_j$ and the corresponding location $(x_j,y_j,f_j)$ \\
			Add the tuple $(x_j,y_j,f_j, fp_j)$ to the radio map;\\
		}
	}%For
	Return the radio map; \\
\end{algorithm}

\section{Experiments and Results}

The proposed methods were evaluated by experiments conducted in an eight-storey office building. The area of each floor is about 4,570 square meters. The test path goes through two floors of this building, and its length is about 362 meters. 

The device we used is a Google Nexus 6 smartphone equipped with WiFi, accelerometer, magnetometer, gyroscope, and barometer. An Android app was developed to collect the  sensor data. A tester walked along the pre-set path with the phone in hand, and clicked on the app to record the RPs to evaluate the location accuracy. The data recorded include the MAC address of visible APs and corresponding RSS, and readings from the accelerometer, gyroscope, compass, and barometer. Table \ref{tab:parameter_setting} gives the values of parameters used in this paper, which are empirically determined. 

\begin{table}[ht]
	\centering
	\caption{\label{tab:parameter_setting} Parameter setting}
	\resizebox{0.4\textwidth}{!}{
	\begin{tabular}{cccc}
		\toprule[0.5pt]
		\textbf{Function} & \textbf{Paramater} & \textbf{Value} \\
		\midrule[0.5pt]
		\multirow{3}{*}{Acc landmark detection} & Window size & 50 samples\\ %\cline{2-3}
		& Walking state threshold & 2 s\\ %\cline{2-3}
		& Still state threshold & 1 - 8 s\\ %\cline{2-3}
		\hline 
		\multirow{2}{*}{Gyro landmark detection} & Window size & 10 samples\\%\cline{2-3}
		& Gyro threshold $\epsilon_{gyro}$ & 1.1 rad/s\\ %\cline{2-3}
			\hline 
		\multirow{2}{*}{Baro landmark detection} & pressure threshold $\epsilon_{baro_1}$	& 0.05 hPa	\\	%\cline{2-3}
		&  pressure threshold $\epsilon_{baro_2}$	& 0.3 hPa \\ %\cline{2-3}
		\hline
	    \multirow{4}{*}{PDR} & Pressure difference $p_d$ & 0.45 hPa \\ %\cline{2-3}
	    &   Heading threshold $\epsilon_{theta}$  & $30^o$\\%\cline{2-3}
	    & Confidence threshold $\epsilon_v$ & 0.25 \\
	    & Initial step length & 0.63 m\\
	    \hline
	    \multirow{3}{*}{Quality Evaluation}  & Step periodicity threshold $T_{max}$ & 1 s \\%\cline{2-3}
	    &		 Step periodicity threshold $T_{min}$ & 0.4 s	\\%\cline{2-3}
	    &   Belief threshold $\epsilon_{S}$ & 15 \\
		\bottomrule[0.5pt]
	\end{tabular}
	}
\end{table}

\subsection{Accuracy Evaluation of Landmark Graph-based Localization}
\begin{figure}[h]
	\centering 
	\includegraphics[width=0.4\textwidth]{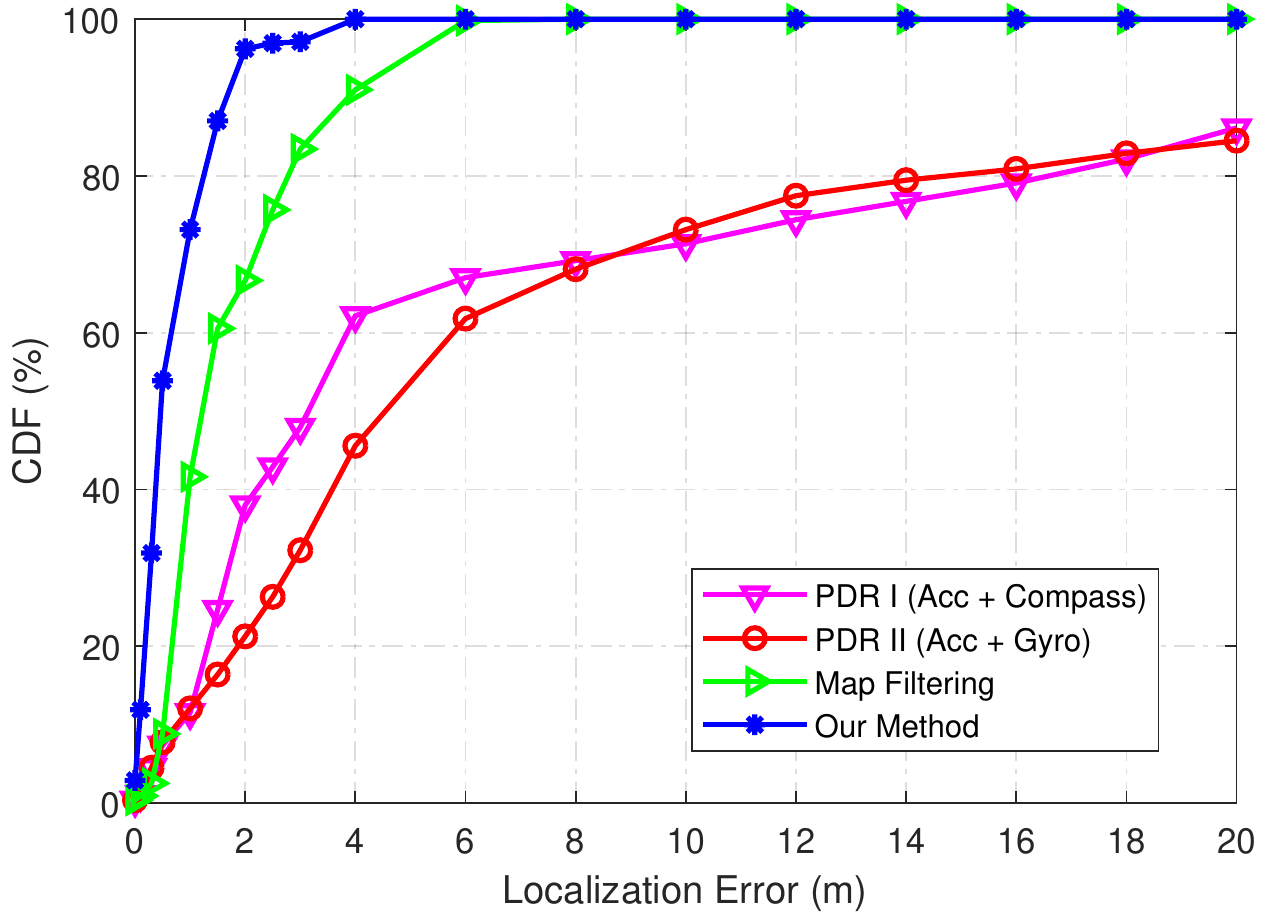}
	\caption{Localization accuracy comparison of different methods used to estimation the location of RPs}
	\label{fig:CDF}
\end{figure}

We first evaluate the accuracy of the landmark graph-based indoor localization method, which is used to obtain the location estimation for associating a fingerprint. To demonstrate the superiority of the landmark graph-based method, we compare it with the commonly-used PDR methods \cite{23} and map filtering method \cite{25}. Both PDR I and PDR II methods use the step-counting-based approach to estimate the step length, however PDR I uses the compass readings to estimate the heading while PDR II uses the gyroscope to calculate the heading. The map filtering method fuses the PDR I with a floor plan with 200 particles. Figure \ref{fig:CDF} shows the cumulative distribution function (CDF) of the localization error of different methods. It shows that our method significantly outperforms the other methods, achieving a mean error of 0.71 meters. 

\subsection{Effect of Quality Control}
\begin{figure}[h]
	\centering 
	\includegraphics[width=0.4\textwidth]{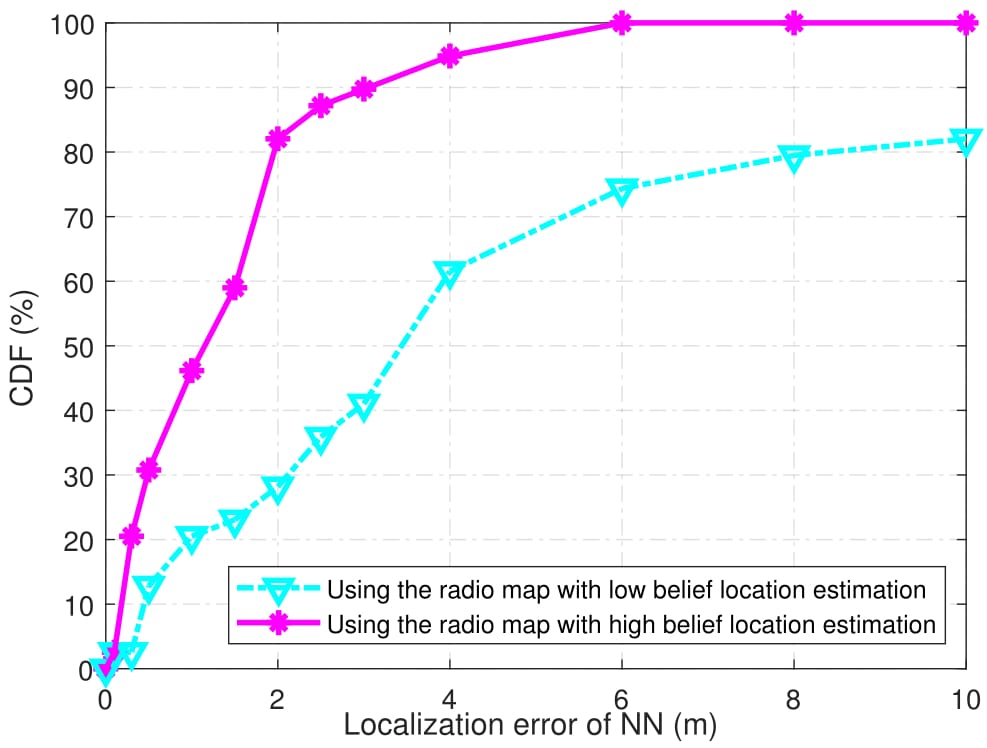}
	\caption{Localization error of 1-NN using the radio map formed with different beliefs}
	\label{fig:result_quality_control_location}
\end{figure}

To evaluate the effect of the accuracy of estimated locations that are used to associate  fingerprints, we conducted two experiments along a straight path segment such that we can ignore the effect of heading. In the first experiment, the user walked normally at a consistent speed; While in the second experiment, the user walked with varying speeds and stopped at a few locations to introduce some false walking noise. The mean error of PDR was less than 1 meter in the first experiment, and about 5 meters in the second experiment. Figure \ref{fig:result_quality_control_location} shows the localization error of 1-NN by using the radio map that was associated with location estimation with different beliefs. It can be seen that using the location estimation with high belief ($bel(S) > 18$) to associate RSS measurements for constructing a radio map achieves much higher accuracy than that with low belief ($bel(S) < 10$). Therefore, it is necessary to control the quality of location estimates that are used to associate fingerprints, which has a direct effect on the accuracy of the fingerprinting localization method. 

\subsection{Fingerprint Collection Method Comparison }

We compare the proposed fingerprint collection method with the state-of-the-art Zee system \cite{14} and the commonly-used manual method, in which the user locates herself on a gird of points on the floor map and collects a number of WiFi scans. For the manual method, we collect fingerprints along the test path with a distance interval of 1.5 meters. At each reference point, 5 WiFi scans and 10 WiFi scans for the manual method were considered. For the Zee system and our method, only one WiFi scan was collected at each reference point. In total, we collected 252 RPs with a distance interval of 1.5 meters for the manual method, and 609 RPs for our method. The locations of RPs for the manual method must be inputted to the system, while those for our method are automatically obtained from the landmark graph-based location estimation method.  

We first compare the time spent by the different methods on collecting fingerprints, including setup time and data collection time. Table \ref{tab:time_comp} shows the approximate time spent by different methods. The setup time of the manual method is about 90 minutes, mainly spent on measuring the locations of RPs. By contrast, the setup time of the Zee system is similar to that of our method, which is about 10 minutes, mainly used to obtain wall information and the landmark graph, respectively. As for the data collection time, the Zee system and our method require much less time to collect fingerprints for constructing a radio map than the manual method. 

\begin{table}[h]
	\centering
	\caption{\label{tab:time_comp} Time spent on fingerprint collection}
	\begin{tabular}{cccc}
		\toprule[0.5pt]
		\textbf{Method} & \textbf{Setup (mins)} & \textbf{Data collection (mins)}\\
		\midrule[0.5pt]
	%	Point-based (1 scan)  & 90 & 30\\		
		Manual method (5 scans) & $\sim$ 90 &$\sim$ 50\\
		Manual method (10 scans) &$\sim$ 90 & $\sim$ 100\\
		Zee & $\sim$ 10 & $\sim$ 10 \\
		\textbf{Proposed method} & \textbf{$\sim$ 10} & \textbf{$\sim$ 10}\\
		\bottomrule[0.5pt]
	\end{tabular}
\end{table}

We compare the localization error of 1-NN by using the radio map constructed by our method, Zee, and manual method. {\color{black} The positive values fingerprint representation was used \cite{26}, namely
	
$$
positive_i(fp) = \left \{  
\begin{array}{llll}
	RSS_i - min, \quad \text{if the $i$-th AP is present } \\
	   \quad  \quad \text{ in the fingerprint $fp$ and $RSS_i \geq \tau$} \\
 \quad	0,  \qquad \quad \text{otherwise}\\
\end{array} 
\right.
\eqno (21)
 $$
where $RSS_i$ is the received signal strength from the $i$-th AP and $\tau$ is a threshold value (APs whose RSS were lower than the threshold are considered as not-detected), and $min$ is the lowest RSS value minus 1 considering all the fingerprints. Two distance metrics, namely Euclidean and Sorensen, were considered. 
\begin{figure}[h]
	\centering 
	\includegraphics[width=0.4\textwidth]{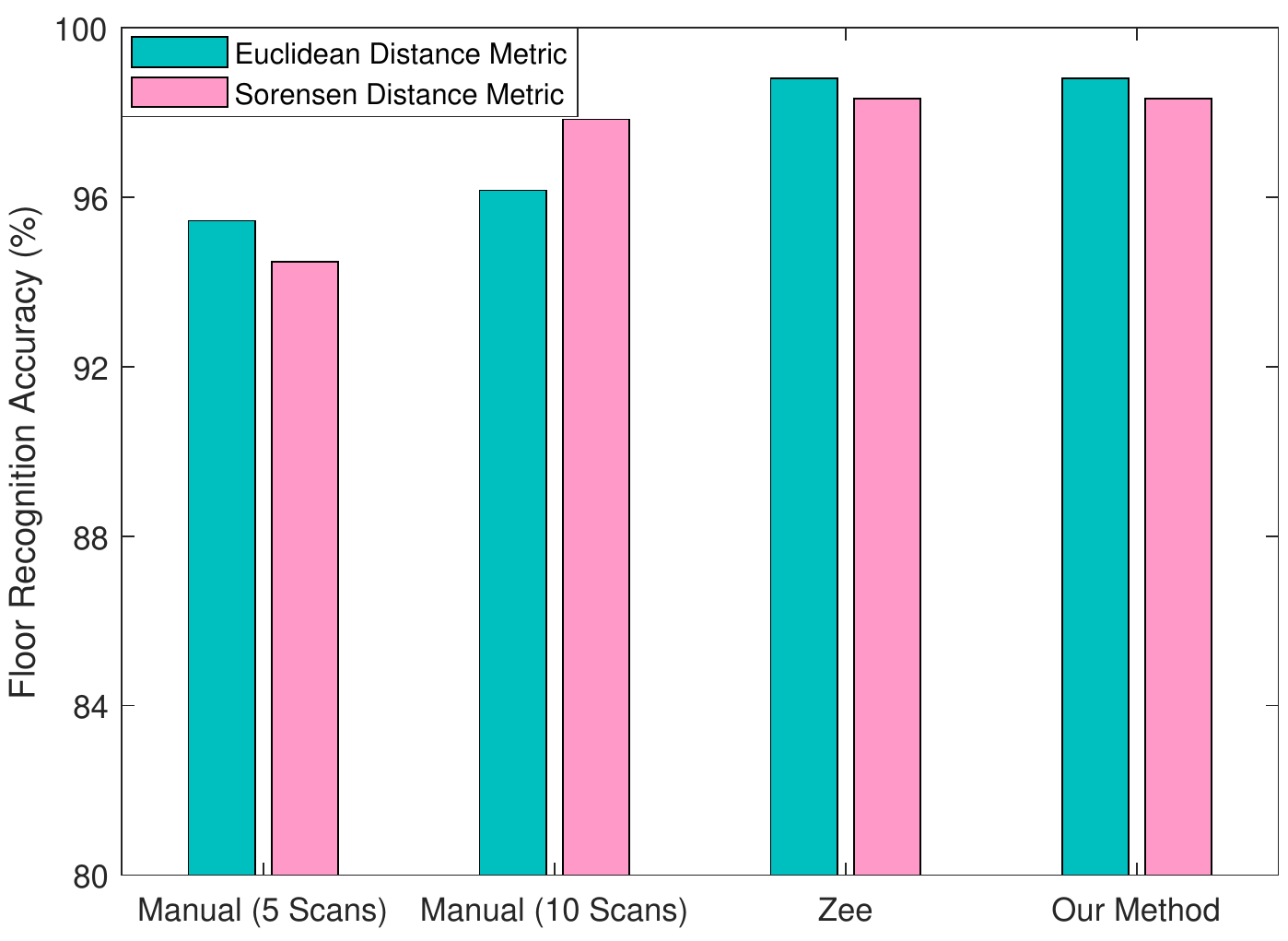}
	\caption{Floor recognition accuracy of 1-NN using the radio map generated by different methods ($\tau$ = -90 dBm)}
	\label{fig:floor_accuracy}
\end{figure}

Figure \ref{fig:floor_accuracy} shows the floor recognition accuracy of 1-NN using the radio map constructed by the different methods. It can be seen that the floor recognition accuracy of our method is similar to that of Zee, which is higher than the manual methods with 5 scans and 10 scans per location point. Using the Euclidean distance metric achieves a bit higher floor recognition rate than that using the Sorensen distance metric. 

\begin{figure}[h]
	\centering 
	\includegraphics[width=0.4\textwidth]{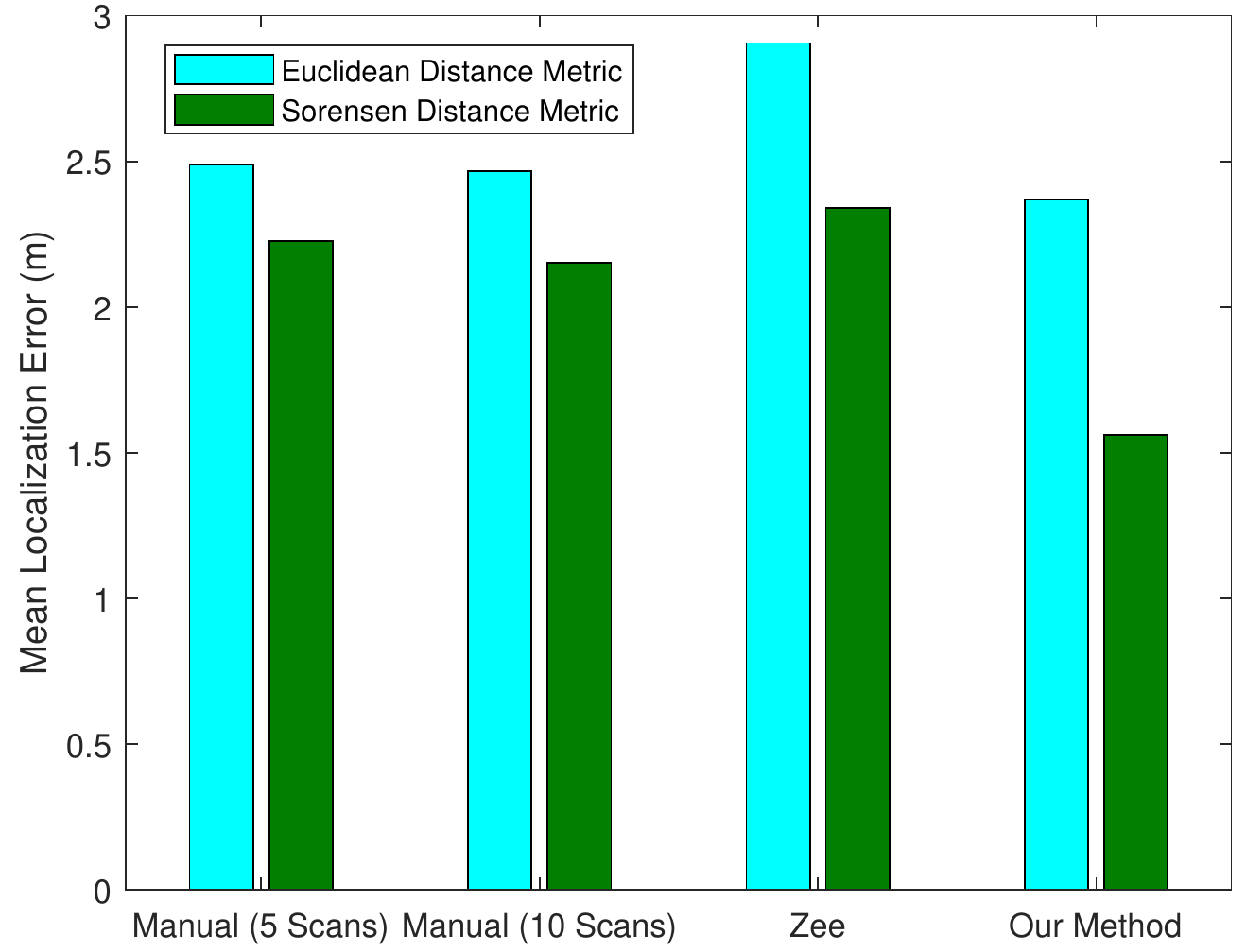}
	\caption{Mean localization error of 1-NN using the radio map generated by different methods ($\tau$ = -90 dBm)}
	\label{fig:mean_error}
\end{figure}
Next, we analyze the localization error of 1-NN using the radio map generated by different methods. Note that the mean localization error was computed when the floor of a location point was correctly recognized. As shown in Figure \ref{fig:mean_error}, our method outperforms the state-of-the-art Zee and the commonly-used manual methods. Our method achieves a mean error of about 1.5 meters, which is considerably higher than Zee (about 2.3 meters) and the manual methods (about 2.2 meters).

\begin{figure}[H]
	\centering 
	\includegraphics[width=0.4\textwidth]{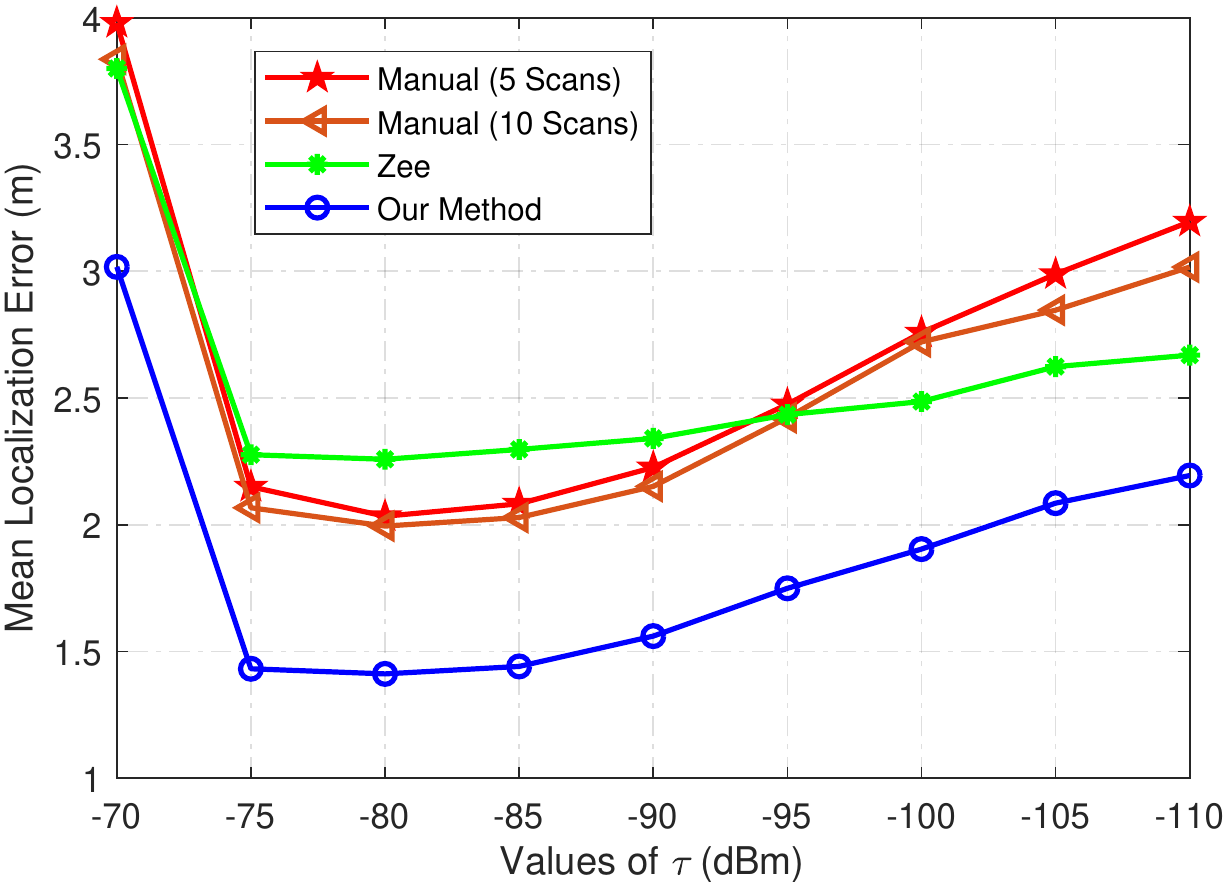}
	\caption{Effect of RSS threshold on the mean localization error of 1-NN (using Sorensen distance metric)}
	\label{fig:effect_of_tau}
\end{figure}

The effect of RSS threshold is demonstrated in Figure \ref{fig:effect_of_tau}. The best performance of all the methods is achieved when the RSS threshold $\tau$ was set to -80 dBm. A larger or smaller value of $\tau$ will lead to an increase in the mean localization error. This is because increasing the RSS threshold may introduce APs with weak signal, which are vulnerable to human movements, and decreasing the RSS threshold would exclude some useful APs that can help improve the localization accuracy.

\section{Conclusion } %and Future Work
In this study, we present a novel fast fingerprint collection method, which aims to reduce the labor and time required to associate RSS measurements with their corresponding locations. The landmark graph-based method is used to automatically estimate the location of reference points, which outperforms the commonly-used PDR method and map filtering method. We also compare the accuracy of radio map constructed by different methods, and experimental results show that the proposed method achieves a better accuracy than manual method and the state-of-the-art Zee system. 

Although the proposed method has significantly reduced the time and labor needed to construct a radio map, it requires the support of spatial constraints. This means that it may fail in open large indoor areas where spatial constraints are insufficient. In the future, we will work on implementing a universal indoor localization solution that integrates not only WiFi fingerprints, magnetic fingerprints and landmarks, but also semantic features and other salient features, which will work in different situations. 

\section*{Acknowledgment}
 This work is supported by the National Key Research and Development Program of China (No. 2016YFB0502200).

%%%%%%%%%%%%%%%%%%%%%%%%%%%%%%%%%%%%%%%%%%
\vspace{6pt}

%%%%%%%%%%%%%%%%%%%%%%%%%%%%%%%%%%%%%%%%%%
%=================================================================
% References: Variant A
%=================================================================
% Back Matter (References and Notes)
%----------------------------------------------------------


\begin{thebibliography}{1}

% Reference 1
\bibitem{1}
A. Tahat, G. Kaddoum, S. Yousefi, S. Valaee, and F. Gagnon, ``A look at the recent wireless positioning techniques with a focus on algorithms for moving receivers," {\em IEEE Access}, vol. 4, pp. 6652-6680, 2016.

\bibitem{2}
S. He and S. H. G. Chan, ``Wi-Fi fingerprint-based indoor positioning: Recent advances and comparisons", {\em IEEE Communications Surveys \& Tutorials}, vol. 18, no. 1, pp. 466-490, 2016.


\bibitem{3}
Y. Zhuang, Z. Syed, Y. Li, and N. El-Sheimy, ``Evaluation of two WiFi positioning systems based on autonomous crowdsourcing of handheld devices for indoor navigation", {\em IEEE Transactions on Mobile Computing}, vol. 15, no. 8, pp. 1982-1995, 2016.

\bibitem{4}
B. Wang, Q. Chen, L. T. Yang, and H. C. Chao, ``Indoor smartphone localization via fingerprint crowdsourcing: Challenges and approaches," {\em IEEE Wireless Communications}, vol. 23, no. 3, pp. 82-89, 2016.

\bibitem{5}
G. Caso, L. De Nardis, F. Lemic, V. Handziski, A. Wolisz, and M. G. Di Benedetto, ``ViFi: virtual fingerprinting WiFi-based indoor positioning via multi-wall multi-floor propagation model," {\em arXiv preprint arXiv }:1611.09335, 2016.

\bibitem{6} % 5-6
B. Ferris, D. Fox, and N. D. Lawrence, ``WiFi-SLAM using Gaussian process latent variable models," in {\em Proc. 20th Int. Joint Conf. Artif. Intell.}, 2007, pp. 2480-485.

\bibitem{7}
J. Huang, D. Millman, M. Quigley, D. Stavens, S. Thrun, and A. Aggarwal, ``Efficient, generalized indoor wifi graphslam," In {\em IEEE International Conference on Robotics and Automation (ICRA)}, 2011, pp. 1038–1043. 

\bibitem{8}
L. Bruno and P. Robertson, ``WiSLAM: Improving FootSLAM with WiFi," in {\em Proc. Int. Conf. Indoor Positioning Indoor Navigat.}, 2011, pp. 1–10.

\bibitem{9}
R. Faragher and R. Harle, ``SmartSLAM—an efficient smartphone indoor positioning system exploiting machine learning and opportunistic sensing," in {\em the 26th Int. Technical Meeting of the Satellite Division of the Institute of Navigation}, 2013, pp. 1-14.

\bibitem{10}
J. G. Park, B. Charrow, D. Curtis, J. Battat, E. Minkov,  J.  Hicks, S. Teller, and J. Ledlie, ``Growing an organic indoor location system," in {\em ACM Proceedings of the 8th international conference on Mobile systems, applications, and services}, 2010, pp. 271-284.

\bibitem{11}
S. Yang, P. Dessai, M. Verma, and M. Gerla, ``FreeLoc: calibration-free crowdsourced indoor localization," In {\em Proc. IEEE INFOCOM)}, 2013, pp. 2481-2489.

\bibitem{12}
J. Ledlie, J. G. Park, D. Curtis, A. Cavalcante, L. Camara, A. Costa, and R. Vieira, ``Mol\'{e}: a scalable, user-generated WiFi positioning engine," {\em Journal of Location Based Services}, vol. 6, no. 2, pp. 55-80, 2012.

\bibitem{13}
K. Chintalapudi, A. P. Iyer, and V. N. Padmanabhan, ``Indoor localization without the pain", In {\em Proceedings of the sixteenth annual international conference on Mobile computing and networking}, 2010, pp. 173-184.

\bibitem{14}
A. Rai, K.K. Chintalapudi, V.N. Padmanabhan, and R. Sen, ``Zee: Zero-effort crowdsourcing for indoor localization,"In {\em Proceedings of the 18th annual international conference on Mobile computing and networking}, 2012, pp. 293-304.

\bibitem{15}
Z. Yang, C. Wu, and Liu, Y, ``Locating in fingerprint space: wireless indoor localization with little human intervention", In {\em Proceedings of the 18th annual international conference on Mobile computing and networking}, 2012, pp. 269-280. 

\bibitem{16}
F. Gu, A. Kealy, K. Khoshelham, and J. Shang, ``Efficient and accurate indoor localization using landmark graphs", {\em International Archives of the Photogrammetry, Remote Sensing \& Spatial Information Sciences}, vol. XLI-B2, pp. 509-514, 2016.

\bibitem{23} % 16-23
H. Wang, S. Sen, A. Elgohary, M. Farid, M. Youssef, and R.R. Choudhury, ``No need to war-drive: Unsupervised indoor localization," In {\em Proceedings of the 10th international conference on Mobile systems, applications, and services}, 2012, pp. 197-210.

\bibitem{24}
F. Gu, K. Khoshelham, J. Shang, F. Yu, and Z. Wei, ``Robust and accurate smartphone-based step counting for indoor localization", {\em IEEE Sensors Journal}, vol. 17, no. 11, pp. 3453-3460, 2017.

\bibitem{25}
S. J. Sadiq and S. Valaee, ``Automatic device-transparent RSS-Based indoor localization", In {\em Global Communications Conference (GLOBECOM), 2015 IEEE}, 2015, pp. 1-6.

\bibitem{26}
X. Hu, J. Shang, F. Gu, and Q. Han, ``Improving Wi-Fi indoor positioning via AP sets similarity and semi-supervised affinity propagation clustering", {\em International Journal of Distributed Sensor Networks}, vol. 2015, no. 109642, pp. 1-11, 2015.


\end{thebibliography}
\end{document}